\documentclass[showpacs,twocolumn,pre]{revtex4-1}
\usepackage{amssymb,amsmath,amsthm,mathtools,color}

\begin{document}
\title{Coarsening and clustering in run-and-tumble dynamics with short-range exclusion}

\author{N\'estor Sep\'ulveda}
\author{Rodrigo Soto}
\affiliation{Departamento de F\'{\i}sica, Facultad de Ciencias F\'{\i}sicas y Matem\'aticas, Universidad de Chile, Avenida Blanco Encalada 2008, Santiago, Chile}

\date{\today}

\begin{abstract}

The emergence of clustering and coarsening in crowded ensembles of self-propelled agents is studied using a lattice model in one dimension. The persistent exclusion process, where particles move at directions that change randomly at a low tumble rate $\alpha$, is extended allowing sites to be occupied by more than one particle, with a maximum $n_\text{max}$ per site. Three phases are distinguished. For $n_\text{max}=1$ a gas of clusters form, with sizes distributed exponentially and no coarsening takes place. For $n_\text{max}\geq 3$ and small values of $\alpha$, coarsening takes place and few large clusters appear, with a large fraction of the total number of particles in them. In the same range of $n_\text{max}$ but for larger values of $\alpha$, a gas phase where a negligible fraction of particles takes part of clusters. Finally, $n_\text{max}=2$ corresponds to a crossover phase. The character of the transitions between phases is studied extending the model to allow $n_\text{max}$ to take real values and jumps to an occupied site are probabilistic. The transition from the gas of clusters to the coarsening phase is continuous and the mass of the large clusters grows continuously when varying the maximum occupancy, and the crossover found corresponds to values close to the transition. The second transition, from the coarsening to the gaseous phase can be either continuous or discontinuous depending on the parameters, with a critical point separating both cases.

\end{abstract}
\maketitle

\section{Introduction\label{sectionIntroduction}}

Collections of self-propelled agents like fish, birds, bacteria, microtubules, or even non-biological objects as active colloids, present the remarkable property that they can gather together in clusters.  The phenomenology of clustering is large, with important differences between, for example, fish and bird,  where inertia and social behavior are important, and microscopic organisms  (see Ref.~\cite{vicsek2012} and references therein).
In the case of microorganisms swimming at low Reynolds numbers, hydrodynamic interactions emerge, which can explain in part their clustering (see Ref.~\cite{Lauga2009} and references therein, for a description of the hydrodynamics of microorganisms). However, in many cases there is no evident attraction between individuals but nevertheless clustering takes place. Different mechanisms have been proposed to explain and quantify this phenomenon. One paradigm is that mutual alignment of the self-propelled individuals, along the line of the Viscek model and others, results in   swarming clusters~\cite{vicsek1995}. Also, motility-induced phase separation (MIPS) has been proposed as a driving mechanism \cite{tailleur2008,cates2015}: the mutual interactions gives rise to an effective reduction of the motility when the local density increases, resulting in a positive feedback for clustering. Finally, it has been shown that the combination of excluded volume and persistence in the direction of motion leads also to the formation of clusters \cite{peruani2006,thompson2011,fily2012,therkauff2012,bialke2013,redner2013,buttinoti2013,palacci2013,soto2014,levis2014,bialke2015,locatelli2015,slowman2016}. Indeed, if particles persist in their direction and encounter others that cannot overlap, they will remain blocked until they change direction, for example, by tumbling or rotational diffusion. This can be understood as an extreme case of MIPS where motility is reduced to zero by exclusion.

The case when clustering appears by the combination of excluded volume and persistence has been studied using off-lattice models, with  agents that change direction continuously by the effect of rotational diffusion \cite{peruani2006,fily2012,therkauff2012,bialke2013,redner2013,buttinoti2013,palacci2013,levis2014,bialke2015,locatelli2015}. Normally, to avoid effects due to alignments, spherical agents are considered. Also, lattice models have been proposed where cells move in prescribed directions, which change randomly at tumble events \cite{thompson2011,soto2014,slowman2016}. Both in the lattice and off-lattice models, in absence of other attractive interaction, it has been observed that clusters can appear (see the recent reviews in Refs.~\cite{bialke2015,cates2015}). Two cases are observed: the steady state consists of many small clusters, which evaporate and merge and actively exchange particles among them, or coarsening takes place and few large clusters remain in the steady state. Ultimately, these clusters will merge into a single one on large time scales. Using as order parameter the fraction of particles in the largest cluster, a clustering to coarsening transition has been identified in terms of the activity for an active colloidal suspension \cite{buttinoti2013}. In a similar system, a phase diagram has been constructed in terms of density and activity, finding gas, clustering, and percolating phases \cite{levis2014}. The authors, however, indicate that going from gas to cluster there is no real phase transition but, rather, a crossover takes place, and no distinction is made between clustering and coarsening. In similar system and control parameters, a phase diagram is presented in Refs.~\cite{redner2013,bialke2013}, where coarsening and gas phases are present. 
The objective of the present work is to study systematically the phase diagram for self-propelled particles that interact with excluded volume, in order to determine whether the gaseous, clustering and coarsening phases are well defined and, in this case, characterize the transitions between them.
For that purpose, we adopt a lattice model, which is the simplest that includes persistence and excluded volume and no residual interactions can appear.

Previous studies show that in the lattice model, clusters form at low tumbling rates,  although with important differences depending if the excluded volume is strictly enforced with maximum one particle per site \cite{soto2014} or if many particles per site are allowed \cite{thompson2011}. In the first case, clusters form with an exponential size distribution, but no coarsening takes place, while in the second case, coarsening happens and the final state consists of few large clusters. This difference in behavior has its origin only  on the maximum number of particles per site, which has been explored to be either 1 or very large, and no intermediate values have been studied. In this manuscript we explore how this parameter affects coarsening and the character of the associated transition. The pair distribution for the case with maximum one particle per site has been found exactly and it has been shown that clusters appear but no coarsening or phase transitions take place \cite{slowman2016}, in agreement with  numerical results \cite{soto2014}.

Our observations show that indeed three distinct phases exist: gaseous, clustering and coarsening. To characterize the transitions the model is extended to allow for the maximum number or particles per site to be a real number and jumps to a partially occupied site are probabilistic. It is shown then that authentic non-equilibrium transitions between phases takes place, that change from continuous to discontinuous at a critical point.

\section{Model\label{sectionModel}}

We consider a periodic lattice in one dimension, consisting of $L$ sites, where $N$ particles move, with time evolving in discrete time steps. Each particle has a state variable $s=\pm 1$ that indicates the direction to which it points. At each time step particles attempt to jump one site in the direction pointed by $s$ and excluded volume effects takes place in the following way: if there are less than $n_\text{max}$ particles in the destination site, the jump is performed, otherwise the particle stays in the original position. To model tumble events, the directions $s$ are changed  to  new random directions with a rate $\alpha$. This model is an extension of the persistent exclusion process (PEP) presented in Ref. \cite{soto2014} to the case where multiple occupation per site is allowed. More explicitly, this model reproduces the PEP when $n_\text{max}=1$. The position update of the particles is asynchronous to avoid two particles attempting to jump to the same site simultaneously. That is, at each time step, the $N$ particles are sorted randomly and, sequentially, each particle jumps to the site pointed by $s$ if the occupation in the destination site is lower than $n_\text{max}$. Finally, at the end of the update phase, tumbles are performed: for each particle, with probability $\alpha$, the director $s$ is redrawn at random, independently of the original value. In analogy to hard-core systems, particles only interact via excluded volume effects and there are no energetic penalizations, partial motility reduction as in MIPS, or explicit alignment as in Ref.~\cite{peruani2011}.

The control parameters are the particle density $\phi=N/L$, the tumble rate $\alpha$ and the maximum occupation number $n_\text{max}$. Time is measured in time steps and lengths in lattice sizes, therefore particles have velocities $0,\pm1$. An important order parameter of the system is the fraction of jammed particles $J$, namely those that could not jump, which equals one minus the average squared velocity. If all particles participate in a single cluster, its size would be $L_c=N/n_\text{max}$, with  $L_c/L=\phi/n_\text{max}$  the fraction of the system occupied by the cluster. Here, $\phi/n_\text{max}$ measures the faction of the maximum occupation in the system.
The model has the intrinsic time scale corresponding to the time it takes a particle to cross one site $t_\text{cross}=1$ and also the mean flight time, which depends on density. Tumbling is characterized by the time scale $t_\text{tumble}=1/\alpha$, which must be compared with the previous two, rendering the phase diagram highly non trivial.

Three initial conditions are used. First, particles are placed randomly, respecting the excluded volume restriction, with random orientations to study how clusters are formed from a gas phase. Second, to analyze the stability of clusters,  particles are placed forming a saturated cluster, with $n_\text{max}$ particles per site, where $L_c$ sites are  occupied. Particles in the left half of the cluster point to the right and the other half point to the left. This configuration is stable for  vanishing tumble rate with all particles in it jammed. The remainder particles of the integer division $N/n_\text{max}$ are placed randomly as a gas. The complementary use of these two initial conditions allows us to identify cases where clusters form by nucleation with long nucleation times and, by comparison of the outcome of the two initial conditions, it is possible to verify that steady state has been achieved. Finally, a third initial condition consists of two nearby clusters, with the objective to test if coarsening takes place. 

In all simulations presented in this article, we have chosen $L=2000$ sites with total simulation times equal to $T=1.3\times 10^7$ time steps. Simulations are long enough to reach steady state and to achieve good statistical sampling.

\section{Qualitative description}

\begin{figure}[htb]
\includegraphics[width=\columnwidth]{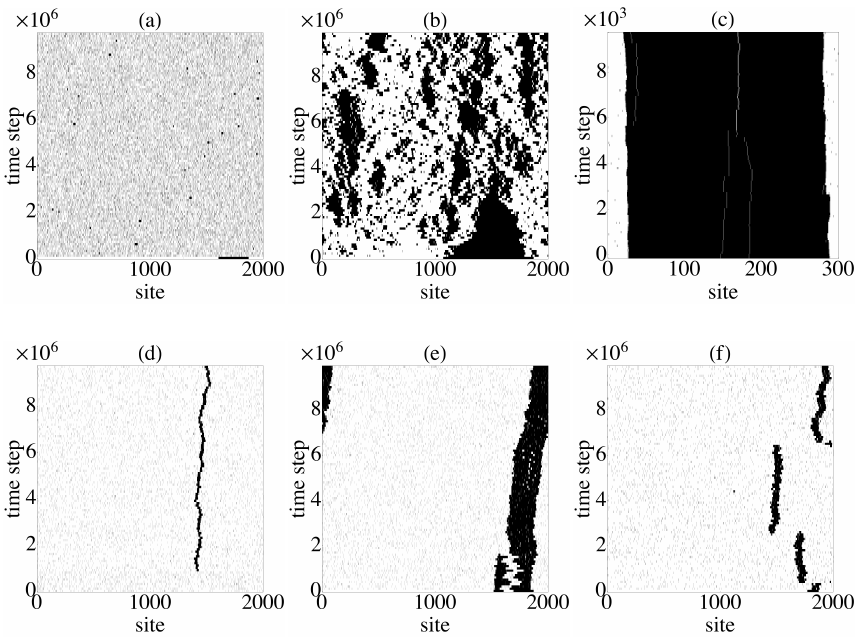}
\caption{Spatio-temporal diagrams for various values of the parameters, with $L=2000$ sites in all cases. 
(a) $n_\text{max}=3$, $\alpha=0.065$, $\phi = 0.4$; (b) $n_\text{max}=1$, $\alpha=0.001$, $\phi = 0.4$; (c) $n_\text{max}=3$, $\alpha=0.002$, $\phi = 0.4$; (d) $n_\text{max}=3$, $\alpha=0.01$, $\phi = 0.1$; (e) $n_\text{max}=3$, $\alpha=0.01$, $\phi = 0.4$; (f) $n_\text{max}=2$, $\alpha=0.01$, $\phi = 0.1$.
The gray level of the dots are proportional to the occupation of a site, with black indicating full occupation. The initial conditions consist of a gas state for (d), one cluster for (a), (b), and (f), and two clusters for (e). 
 On figure (c), the initial condition corresponds to a steady state solution, in a case where a single cluster forms, and the spatial and temporal scales have been changed to highlight the defect dynamics inside the cluster.
}
\label{qualitativeDescription}
\end{figure}

Figure \ref{qualitativeDescription} shows the spatiotemporal diagrams obtained in simulations of the model for different parameters and the three initial conditions  previously described.
Three final states are observed. Particles can form a gaseous phase, with few particles jammed while the majority moves freely. This phase is achieved for large values of $n_\text{max}$ and values of $\alpha$ larger than  a certain threshold that depends on $n_\text{max}$ [Fig.~\ref{qualitativeDescription}(a)]. In the second phase, particles can jam in many clusters of various sizes, giving rise to a gas of clusters phase [Fig.~\ref{qualitativeDescription}(b)]. This phase was previously obtained for $n_\text{max}=1$ \cite{soto2014}, where cluster sizes are distributed exponentially. Finally, in the coarsening phase, particles are jammed in few clusters  of macroscopic size [Fig. \ref{qualitativeDescription}(c)]. Here, the total number of particles in single large clusters constitute an important fraction of the total number of particles. This phase is obtained for low values of tumbling rate when $n_\text{max}>1$, consistent with a related model presented in Ref.~\cite{thompson2011}.

In the cases where the final state presents no large clusters, if a large cluster is initially seeded, it evaporates as shown in Figs.~\ref{qualitativeDescription}(a) and \ref{qualitativeDescription}(b). On the other cases, when large clusters are present in the final state, they are started spontaneously as a result of nucleation from the gas phase, as it is shown in Fig.~\ref{qualitativeDescription}(d), or the initially seeded cluster adjusts its size, evaporating particles, to the final state.
For $n_\text{max}=1$, large clusters are not stable no matter how small we take $\alpha$, in agreement with previous observations \cite{soto2014,slowman2016}, and the initially seeded cluster evaporates to finally give rise to a gas of clusters. 
For values of $n_\text{max}>1$, clusters are stable for small values of the tumbling rate but evaporate when increasing this value.

In the coarsening phase, there is a tendency to form large clusters. This process is extremely slow, as particles in a cluster cannot move, and clusters move as a result of the effective diffusion of the borders by absorption--evaporation processes. Figure~\ref{qualitativeDescription}(e) displays the merging of two clusters, seeded initially close to each other. This processes is conducted by the evaporation of one cluster into the second, by random tumbling of the particles in their borders. As a result of the accumulated random processes, there is a larger probability for the smaller cluster to evaporate into the larger, hence leading to a coarsening process.
Fluctuations can also produce the complete evaporation of a cluster into the gas, increasing its density beyond the saturation point; as a result a cluster is later nucleated as shown in Fig.~\ref{qualitativeDescription}(f).

\begin{figure}[htb]
\includegraphics[width=\columnwidth]{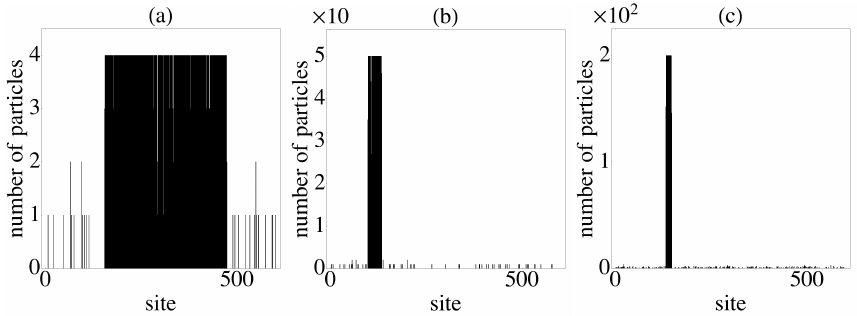}
\caption{Instantaneous snapshot of clusters, displaying the number of particles per site. In all cases the system is in the steady state and a single cluster has formed. 
(a) $n_\text{max}=4$, $\alpha=0.01$, $\phi = 0.7$; (b) $n_\text{max}=50$, $\alpha=0.001$, $\phi = 1.0$; (c) $n_\text{max}=200$, $\alpha=0.001$, $\phi = 2.0$.
The number of sites is $L=2000$, but only a fraction of the system is presented to show the internal cluster structure and compare the structure of clusters for different values of $n_\text{max}$.}
\label{profiles}
\end{figure}

The clusters in the steady state are saturated with almost all sites with $n_\text{max}$ particles as shown in Fig.~\ref{profiles}. Stability of clusters require that the particles in the border must point to the interior, otherwise they could evaporate immediately. The site at the border is fully occupied. The site beside it, toward the exterior of the cluster, can be partially occupied, again with all particles pointing to the interior. The average occupation of this lateral site, the ``shoulder,'' is $n_\text{max}/2$. In summary, each cluster border consists of two sites, with total occupation $3n_\text{max}/2$.
As a consequence of the exclusion volume, particles inside a cluster cannot move and the cluster borders move by absorption of particles that arrive or by the evaporation of one particle in the border region after tumbling. 
Only the two extreme sites at each border are dynamically active.  The border moves when the last site saturates and a new shoulder site starts to develop or when all the particles in the border evaporate. The interior sites, on the other side, are almost dynamically inert because most of the them are jammed and tumbling does not allow them to escape. A few sites, however, are not fully occupied and defects diffuse in the interior as shown in Fig.~\ref{qualitativeDescription}(c).

In the coarsening phase, where there is a single cluster, the gas has a fixed average density just below clusterization. Hence, the position of the cluster borders, which move by random evaporation and absorption processes, must fluctuate coherently to maintain the total mass of the cluster fixed on average [see Figs.~\ref{qualitativeDescription}(d)--\ref{qualitativeDescription}(f)]

In summary, the global dynamics is characterized by the permanent particle exchange between clusters via partial or total evaporation and the spontaneous nucleation of clusters in gas due to fluctuations.

A common feature is that clusters have abrupt walls as is shown in Fig.~\ref{profiles}. Therefore, 
clusters with a smooth density profile are not stable: before saturation particles are not affected by the presence of other particles in the site and the smooth cluster will evolve either to evaporation or to the formation of a saturated cluster with abrupt walls. 
This feature makes continuous models for motility-induced phase separation, which assume that particle density is a locally smooth field,  unsuitable to describe this system \cite{tailleur2008, thompson2011,bialke2013}.

\section{Clusterization and coarsening}
 
The case with $n_\text{max}=1$, with a gas of clusters of sizes distributed exponentially \cite{soto2014} seems radically different from the cases with large $n_\text{max}$, where few large clusters are formed  and coarsening takes place. With the objective of characterizing the transition between these regimes, we  define and analyze in the steady state two order parameters that give complementary information on the degree of clusterization and coarsening. 

\begin{figure}
\includegraphics[width=\columnwidth]{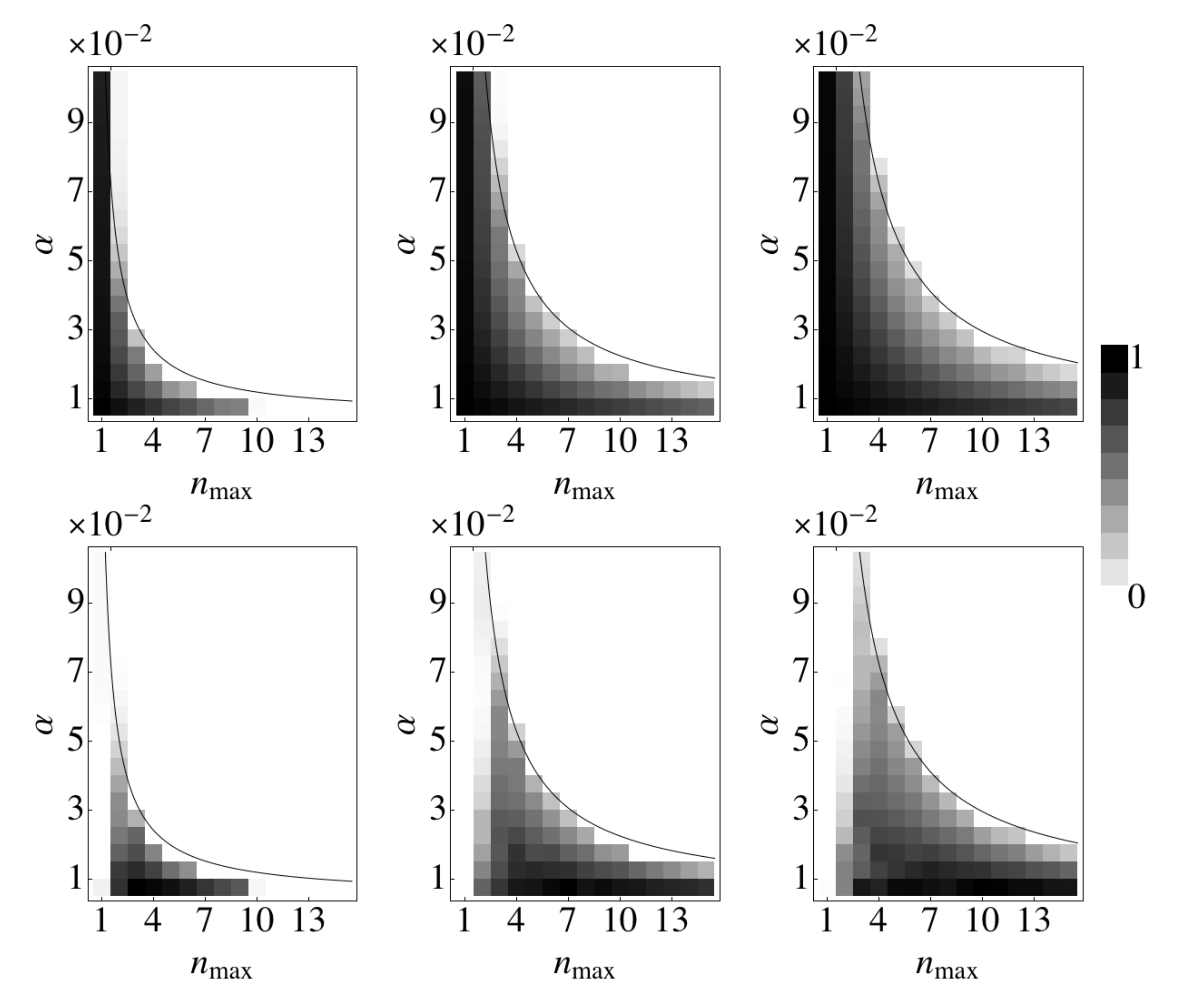}
\caption{
Order parameters $J$ (upper row) and $M$ (lower row) in the parameter space $n_\text{max}$--\,$\alpha$  for $\phi=0.2$ (first column), $\phi=0.5$ (second column),  and $\phi=0.7$ (third column). The continuous black curves correspond to the prediction $\alpha=\frac{\phi}{3\,n_\text{max}}$.
}
\label{orderParameters}
\end{figure}

\begin{figure*}[!htb]
\includegraphics[width=2\columnwidth]{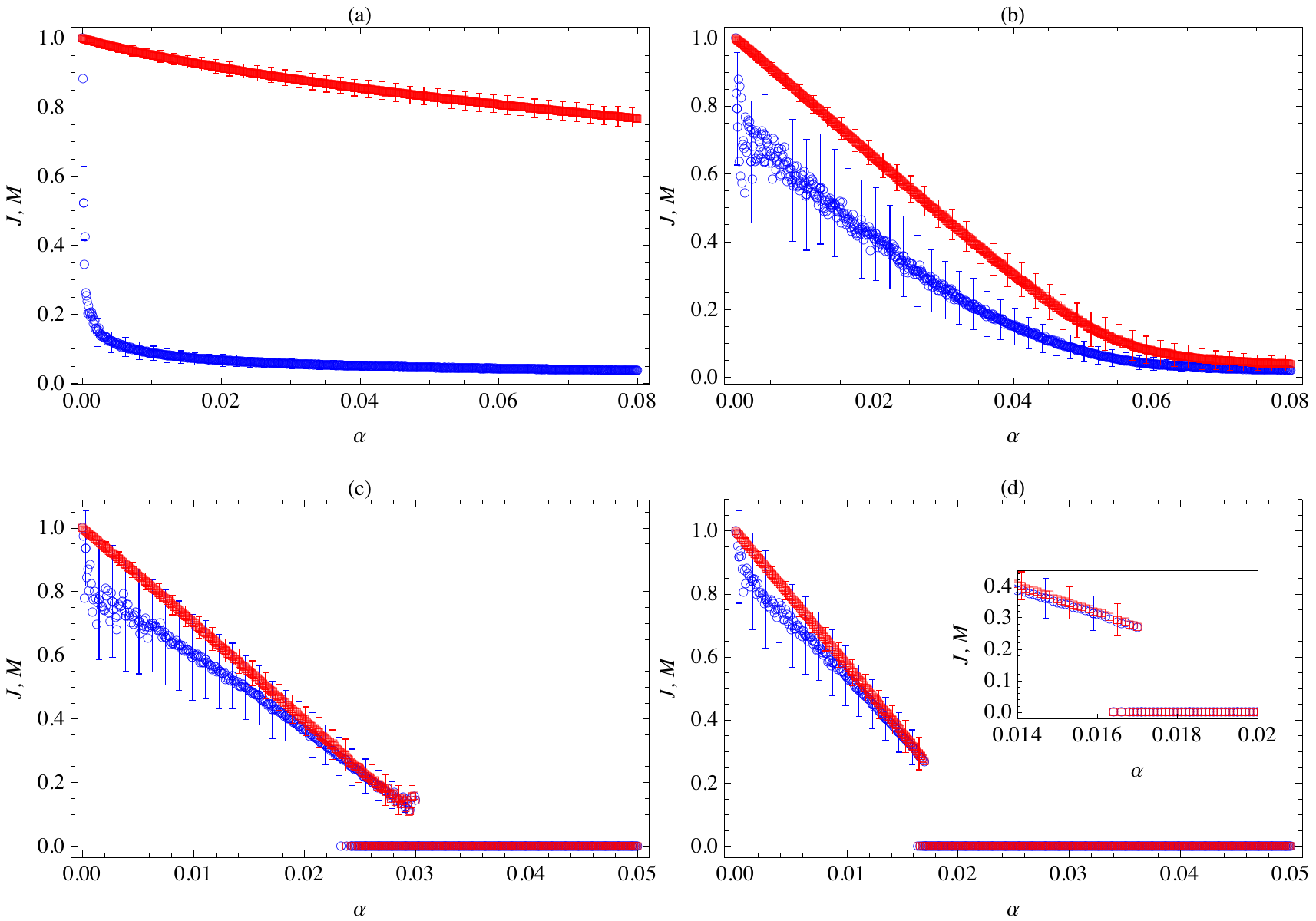}
\caption{ 
(Color online) Order parameters $J$ ($\textcolor{red}{\square}$, red, light gray) and $M$ ($\textcolor{blue}{\bigcirc}$, blue, dark gray) as a function of $\alpha$, for (a) $n_\text{max}=1$, (b) $n_\text{max}=2$, (c) $n_\text{max}=3$, (d) $n_\text{max}=4$, and $\phi=0.2$. The inset in (d) shows the same data in a restricted region of $\alpha$ to highlight the coexistence of the two phases. 
Error bars (evenly shown not to overload the plot) are obtained after $100$ measurements made every $10^5$ time steps, starting from $t=3\times10^6$.}
\label{orderParametersCuts}
\end{figure*}

The first order parameter $J$ is the average fraction of jammed particles, namely those that cannot move because they point to a site that is fully occupied. This parameter measures the number of particles that participate in clusters, independent of the cluster-size distribution.
The upper row of Fig.~\ref{orderParameters} presents $J$ as a function of the control parameters $n_\text{max}$ and $\alpha$, for three values of $\phi$. Clusterization increases by decreasing $\alpha$ and $n_\text{max}$ and it is possible to identify a transition line that separates between clustered and non-clustered regimes, which is more clearly seen in cuts at constant $n_\text{max}$, for $\phi=0.2$, shown by the $\square$ markers in Fig.~\ref{orderParametersCuts}.

The transition line can be computed by balancing the particle fluxes at the cluster boundaries. The outgoing flux of evaporating particles is given by the average number of particles in the two sites that compose each border, $3n_\text{max}/2$, times the tumble rate, resulting in $J_\text{out} = \frac{3}{2}n_\text{max} \alpha$. The incoming flux is given by the particles that approach the boundary, which then equals half the particle density in the gas phase, $J_\text{in}=n_\text{gas}/2$. If $\ell$ is the width of the cluster, the conservation of the total number of particles gives, $n_\text{max}\ell+n_\text{gas}(L-\ell)=L\phi$, which combined with the flux balance results in $\ell=\frac{L(\phi-3\,n_\text{max}\alpha)}{n_\text{max}(1-3\,\alpha)}$. Clusters disappear when $\ell$ vanishes, therefore the transition line is predicted to be at $\alpha=\frac{\phi}{3\,n_\text{max}}$, represented by the black curve in Fig.~\ref{orderParameters}.

 
To identify if particles condense into large clusters, compatible with a coarsening process, we define a second order parameter, related to the size of the clusters. Similarly to Refs.~\cite{buttinoti2013,bialke2013}, we consider  $M$, equal to the average mass of the biggest continuous cluster in the system. By continuous cluster we understand that it has no holes of null occupancy inside it. Coarsening corresponds to most of the jammed particles belonging to a single large cluster, a situation that is determined by the condition $M\approx J$.
The lower row of Fig.~\ref{orderParameters} presents $M$ as a function of the control parameters, $n_\text{max}$ and $\alpha$, for different values of the volume fraction $\phi$, while the $\circ$ markers of Fig.~\ref{orderParametersCuts} present cuts at constant values of $n_\text{max}$ for $\phi=0.2$.

For large $n_\text{max}$, $M$ and $J$ follow similar trends. For small $\alpha$ both take large values, indicating that a large proportion of particles belong to few large clusters, and there is a threshold value of $\alpha$ (depending on $n_\text{max}$) after which both vanish in a discontinuous transition with hysteresis.
For small $n_\text{max}$ the situation is  quite different, where $M$ shows small values even when $J$ is large, indicating that clusters do not present large sizes. That is, there is a region in the phase space, where there is a large proportion of particles belonging to clusters, but these do not coalesce into few large ones. 
The system then presents three phases shown in the left panel of Fig.~\ref{phaseDiagram}: (i) for large values of $n_\text{max}$ and $\alpha$ most of the particles are not jammed, and the system behaves like a gas; (ii) for $n_\text{max}=1$, a significant fraction of the  particles are jammed in small clusters, therefore constituting a gas of clusters; (iii) finally, for small $\alpha$ and  $n_\text{max}>2$ clusters coarsen into a single large cluster. The case $n_\text{max}=2$ is marginal and there is a continuous crossover between phases.

Notably, however, these phases do not respect the level rule of thermodynamics. Indeed, as shown in Fig.~\ref{orderParameters}, the region where a single cluster exists depends on the global volume fraction $\phi$. That is, increasing $\phi$ not only makes clusters bigger but also the relevant region in parameter space gets larger.

\section{Fractional occupation}

\begin{figure*}[!htb]
\includegraphics[width=2\columnwidth]{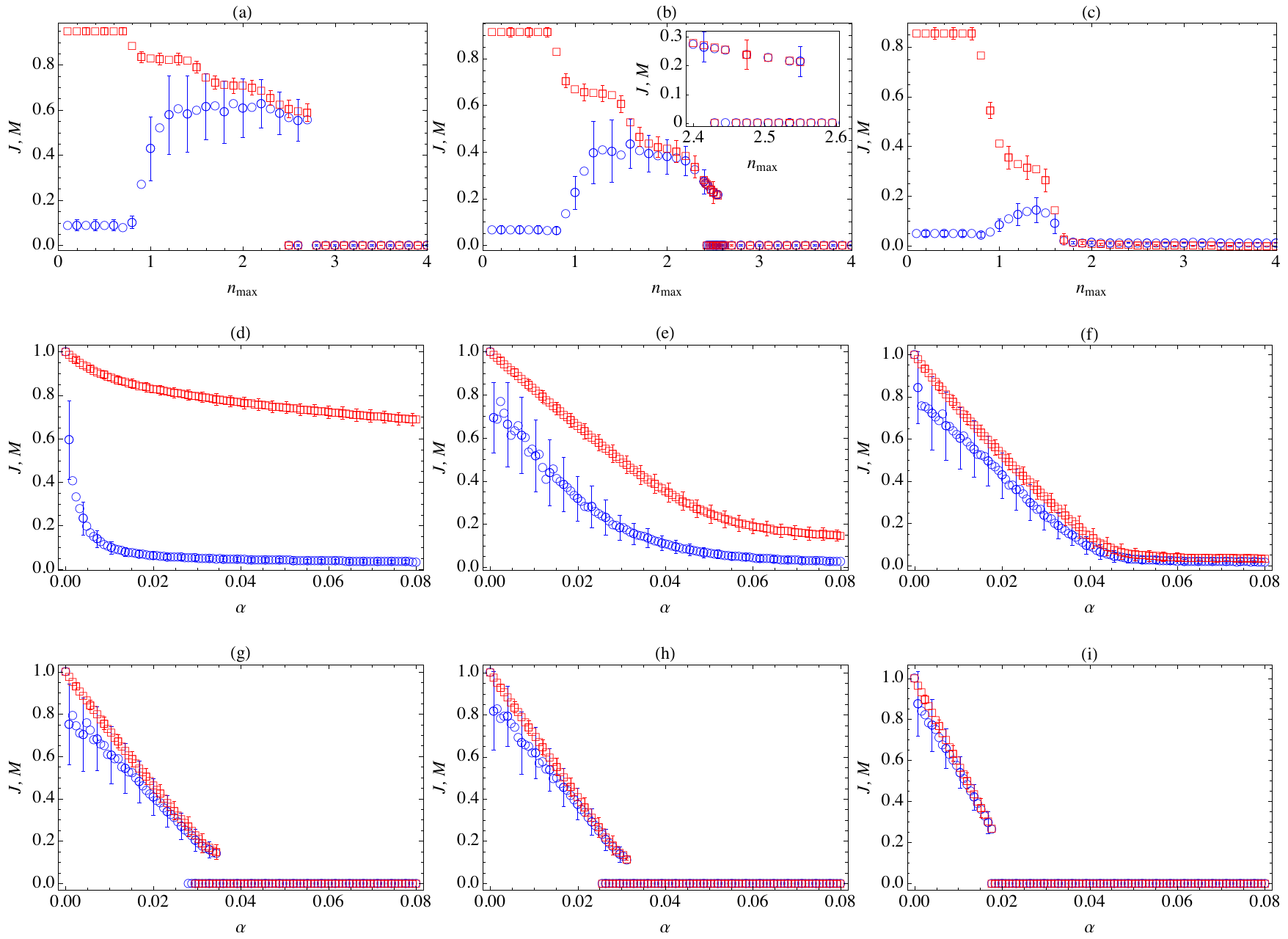}
\caption{(Color online) Order parameters  $J$ ($\textcolor{red}{\square}$, red, light gray) and $M$ ($\textcolor{blue}{\bigcirc}$, blue, dark gray) for the model with fractional occupation, Eq.~(\ref{continuousModel}). 
(a) $\alpha=0.01$,  (b) $\alpha=0.02$, (c) $\alpha=0.04$, (d) $n_\text{max}=0.8$, (e) $n_\text{max}=1.1$, (f) $n_\text{max}=1.6$, (g) $n_\text{max}=1.7$, (h) $n_\text{max}=2.0$ and (i) $n_\text{max}=2.8$.
In all cases, the number of sites is $L=2000$ and the volume fraction is $\phi=0.2$. The upper row shows the order parameters as function of $n_\text{max}$ for fixed $\alpha$, while in the two bottom rows the order parameters are plotted against $\alpha$ for fixed $n_\text{max}$. The inset in (b) shows the same data in a restricted region of $n_\text{max}$ to highlight the coexistence of the two phases. The error bars are obtained with the same prescription as described in the caption of Fig.~\ref{orderParametersCuts}.}\label{orderParCutsContinousModel}
\end{figure*}

Figure \ref{orderParametersCuts} shows that there is a qualitative difference between $n_\text{max}= 1$, 2, and 3. The order parameters evidence a transition only for $n_\text{max}\geq 3$, while they behave analytically for $n_\text{max}\leq 2$. Also, there is a strong difference on the value of $M$ between  $n_\text{max}=1$ and 2.
The difference takes place for such small values of $n_\text{max}$ that, by the discrete nature of this parameter, we cannot  study how these transitions appear when varying at will this control parameter. We extend, therefore, the model. Now $n_\text{max}$ can take fractional  values and instead on taking a binary decision depending of whether the occupation of the  destination site, $n_{i+s}$, is smaller or not than $n_\text{max}$, jumps now take place with a probability that smooths the unit step function used in the previous section,
\begin{equation}
P_{i\rightarrow i+s} = \exp\left[-\left( \frac{ n_{i+s} }{ n_\text{max} }\right)^6\right],
  \label{continuousModel}
\end{equation}
where we recall that $s=\pm 1$ is the director of each particle. The sixth power has been chosen such that the allowed occupancies of a site do not exceed for too much the specified value $n_\text{max}$. Larger powers produce jump probabilities that are too sharp and the model is equivalent to having integer $n_\text{max}$. For smaller powers, the jump probability is extremely smooth and actual occupations can exceed notoriously $n_\text{max}$. Short tests made with other exponents close to 6 give similar qualitative results.

Simulations are run for this model with fractional occupation in the range $0< n_\text{max}\leq 4$  and $0\leq\alpha\leq 0.08$, while the number of sites and global density has been fixed to $L=2000$ and $\phi=0.2$.
A similar phenomenology to the integer model is found for the steady state. For small $n_\text{max}$, the two order parameters decouple, signaling that a gas of clusters exist. For intermediate values, the two order parameters take finite values, indicating that coarsening takes place. Finally, for large values of $n_\text{max}$, clusters do not exist and a gas takes over. Now that the control parameter takes fractional values, it is possible to analyze the character of the transitions between these three phases.

Figure \ref{orderParCutsContinousModel} presents the measured order parameters $J$ and $M$  for the model with fractional occupation against either $n_\text{max}$ or $\alpha$, while the other control parameter is kept fixed to $\phi=0.2$. When the order parameters are plotted against $n_\text{max}$, it is possible to distinguish terraces, which are a result of the actual occupancy in each site being discrete.

The transition from a gas of clusters to the coarsening phase is always continuous and takes place at $n_\text{max}\approx 0.8$, almost independent on the value of $\alpha$. In the gas of clusters phase, $J$ is close to one and $M$ is close to zero, with values that are independent on $n_\text{max}$. After the transition to the coarsening phase, $J$ decreases monotonically, while $M$ increases until both order parameters take similar values, when they start to decrease until the gas phase is reached [see Figs.~\ref{orderParCutsContinousModel}(a)--\ref{orderParCutsContinousModel}(c)]. The continuous change in the order parameters is consistent with the phase crossover observed for $n_\text{max}=2$ in the discrete model. Indeed, in the fractional occupation model, this crossover corresponds to regions of parameters in the coarsening phase where $M$ is still small, but already a large cluster exists although coexisting with other smaller clusters. The fraction of the mass allocated in this large cluster grows when increasing $n_\text{max}$ as shown in Figs.~\ref{orderParCutsContinousModel}(a-c).

The second transition, from the coarsening phase to the gaseous phase can be continuous for small values of $n_\text{max}$ while it becomes discontinuous, presenting hysteresis, for larger values of the maximum occupancy [see Figs.~\ref{orderParCutsContinousModel}(d)--\ref{orderParCutsContinousModel}(i)]. There is a critical point, located roughly at $n^\text{c}_\text{max}\approx 1.65$ and $\alpha^\text{c}\approx0.035$, that separates both types of transitions. A schematic representation of the phase diagram is presented in the right panel of Fig.~\ref{phaseDiagram}.

\section{Conclusion}

The persistent motion of active matter combined by excluded volume effects leads to the formation of aggregates or clusters. In this article, using a lattice model, we have shown that depending on the control parameters, the tumbling rate and the maximum occupancy per site, three phases exist:  gas of clusters, coarsening, and gas. In the first two phases, a macroscopic fraction of the particles belong to clusters, but only in the coarsening phase do few clusters concentrate an important fraction of the particles. In the gas phase, a negligible fraction of the particles participate in clusters. Notably, coarsening can take place only as a result of exclusion and persistence, without the need of other attraction mechanisms, making it possible that these clusters act as seeds  for the early stages of biofilm formation, for example. 

\begin{figure}[htb]
\includegraphics[width=\columnwidth]{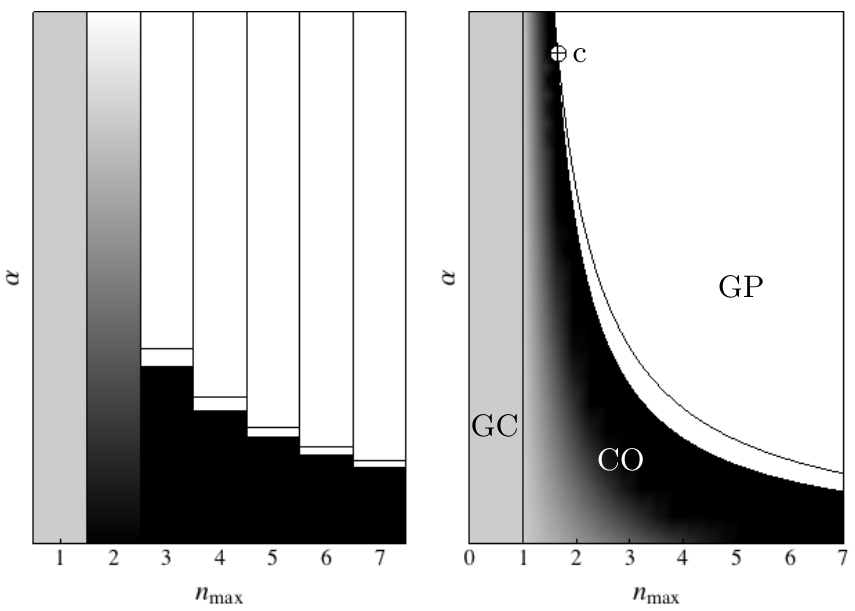}
\caption{Schematic representation of the phase diagrams for the integer (left) and fractional occupation model (right). Black corresponds to the coarsening phase (CO), with few large clusters  that concentrate most of the clustered particles. The clustering or gas of clusters phase (GC) is presented in gray and the gas of particles phase (GP) is in white. The continuous grading of gray, indicates the continuous transition from GC to CO, where the fraction of clustered particles in the large clusters change from negligible to total. For $n_\text{max}=2$ there are no transitions but a continuous crossover between phases in the integer model. The solid curve in the right panel indicates that there is hysteresis in a discontinuous transition, which ends at a critical point (c) signaled by the symbol $\oplus$. In the left panel, the horizontal lines indicate hysteresis.
}
\label{phaseDiagram}
\end{figure}

The extension of the model to allow for the maximum occupancy being a real number and  exclusion being probabilistic, makes it  possible to analyze the character of the transitions between phases. The transition from the gas of clusters to the coarsening phase is continuous, where the total mass in the large clusters grows when varying the maximum occupancy, until a macroscopic fraction of the clustered particles belong to the large clusters. In the coarsening phase, small clusters are always present, although the fraction of particles in them decreases at the expense of large clusters. The second transition, from the coarsening phase to the gas phase can be either continuous or  discontinuous, with a critical point separating these cases. All these  findings conciliates different results in the literature where the three phases were found. These results are summarized in Fig.~\ref{phaseDiagram}, which presents a schematic representation of the phase diagram for the integer and fractional occupation  models.

It remains to be studied how the control parameters in this simple model, can be mapped to other models, for example when particles move off-lattice and the director can change continuously as a result of rotational diffusion. In particular, it will be interesting to analyze if the qualitative phase diagram is preserved.

\acknowledgments
We thank F. Peruani  for fruitful discussions. This research was supported by Fondecyt Grants No.\ 1151029 (N.S.) and 1140778 (R.S.).


\end{document}